\documentclass[aps,prd,twocolumn,superscriptaddress,preprintnumbers,showpacs]{revtex4-1}
\usepackage{amsmath,amsfonts,amssymb,graphicx,color,times,bbm,psfrag,dsfont,
slashed,bigints}
\usepackage[colorlinks=true,citecolor=blue]{hyperref}

\newcommand{\be}{\begin{equation}}
\newcommand{\ee}{\end{equation}}

\newcommand{\bea}{\begin{eqnarray}}
\newcommand{\eea}{\end{eqnarray}}

\newcommand{\hf}{\hat{ \varphi}}
\newcommand{\hvf}{\hat{\bm \varphi}}

\newcommand{\beq}{\begin{equation}}
\newcommand{\eeq}{\end{equation}}

\begin{document}

\title{Multicomponent meson superfluids in chiral perturbation theory}
\author{L. Lepori}
\email{llepori81@gmail.com}
\affiliation{{ Istituto Italiano di Tecnologia, Graphene Labs, Via Morego 30, I-16163 Genova, Italy.}}
\affiliation{Dipartimento di Scienze Fisiche e Chimiche, Universit\`a dell'Aquila, via Vetoio,
I-67010 Coppito-L'Aquila, Italy.}
\affiliation{INFN, Laboratori Nazionali del Gran Sasso, Via G. Acitelli,
22, I-67100 Assergi (AQ), Italy.}
\author{M. Mannarelli}
\email[correspondence at: ]{massimo@lngs.infn.it}
\affiliation{INFN, Laboratori Nazionali del Gran Sasso, Via G. Acitelli,
22, I-67100 Assergi (AQ), Italy.}

\begin{abstract}
We show that the multicomponent meson systems  can be described by  chiral perturbation theory.  We chiefly focus on a system of two  pion gases at different {\color{black} isospin chemical potential},  deriving the general expression of the chiral Lagrangian,  the ground state properties and the spectrum of the low-energy excitations. We consider two different kinds of interactions between the two meson gases: one which does not lock the two chiral symmetry groups and one which does lock them. The former is a kind of interaction that has already been discussed in mutlicomponent superfluids. The latter is perhaps more interesting, because seems to be related to  an instability. Although the pressure of the system does not show any instability, we find that for sufficiently strong locking,  the spectrum of one Bogolyubov mode becomes  tachyonic. This  unstable branch seems to  indicate a  transition to an inhomogeneous phase. 
\end{abstract}
\pacs{}
\maketitle


\section{Introduction} 
Cold hadronic matter   is an interesting playground for a deep understanding of the properties of the strong interaction.  At asymptotic baryonic densities the liberated quarks~\cite{Cabibbo:1975ig} should pair forming a color superconductor, see~\cite{Rajagopal:2000wf,Alford:2007xm,Anglani:2013gfu} for reviews. At  large isospin densities a different kind of  collective phenomenon  happens, with mesons forming a Bose-Einstein condensate (BEC)~\cite{Migdal:1971cu,Sawyer:1972cq,Scalapino:1972fu, Migdal:1973zm,1972PhLA...41..129K,Migdal:1990vm}. In general, the matter density of the system is controlled by the baryonic chemical potential, $\mu_B$, while the isospin chemical potential, $\mu_I$, is associated to its  degree of isospin asymmetry, {\it e. g.} indicating that the number of neutrons differs from the number of protons.
The properties of matter as a function of  $\mu_I$ have  been the subject of intensive investigation for a number of reasons. Systems with large isospin asymmetry exist in Nature; in particular neutron stars~\cite{Shapiro:1983du} are believed to be compact stellar objects  with a large isospin asymmetry. Recently, the possible existence of pion stars has also been proposed~\cite{Carignano:2016lxe, Brandt:2018bwq, Andersen:2018nzq}. 
Regarding the microscopic properties of matter,
 the inclusion of $\mu_I$ can lead to a better understanding of  quantum chromodynamics (QCD) in a regime in which lattice QCD simulation are doable~\cite{Alford:1998sd,Kogut:2002zg,Detmold:2012wc,Detmold:2008yn,Cea:2012ev,Endrodi:2014lja,Brandt:2017oyy,Brandt:2018wkp}.  Remarkably, the lattice QCD simulations of meson gases with vanishing baryonic density are not affected by the sign problem and can be  implemented  for not too high values of $\mu_I$. These simulations are  steadily improving~\cite{Endrodi:2014lja,Brandt:2017oyy,Brandt:2018wkp},  reaching increasingly precise results on the thermodynamic properties of the system and thus offering  powerful tests for the  methods  and models developed for the effective description of the strong interaction. 
 
 Among the various proposed  models, it is worth mentioning  the Nambu-Jona Lasinio (NJL) model~\cite{Nambu:1961tp, Klevansky:1992qe,Buballa:2003qv,Toublan:2003tt,Barducci:2004tt,Barducci:2004nc,Ebert:2005wr,Ebert:2005cs,He:2005nk,He:2010nb} and the quark-meson model~\cite{Klevansky:1992qe, Andersen:2014xxa, Adhikari:2016eef, Adhikari:2018cea}, which can be used in a wide range of values of $\mu_I$. Although  these models are useful tools for exploring the properties of hadronic matter, they  are based on a number of parameters that have to be phenomenologically fixed. Thus, they lead to results which depend on the choice of these parameters and on the number of degrees of freedom used.  Moreover, the obtained results cannot be systematically improved because no expansion parameter can be identified.

A systematic analysis of hadronic matter can be obtained by effective field theories~\cite{Weinberg:1978kz, Pich:1998xt, Holstein:2000ap}, which are based on an   expansion in a control parameter. Here we  focus on  chiral perturbation theory ($\chi$PT), which is an effective theory designed to  describe the low-energy properties of  QCD~\cite{Georgi:1985kw,Leutwyler:1993iq,Gasser:1983yg, Scherer:2002tk,Ecker:1994gg}. The   $\chi$PT Lagrangian  is derived  by the global symmetries of QCD, basically  integrating out the high-energy part.
The effect of the  {isospin chemical} potential is  conveniently included in covariant derivatives, see~\cite{Gasser:1983yg, Leutwyler:1993iq,Leutwyler:1996er} for a general discussion. 
This  approach leads to systematic results, which can be improved including higher orders in the  $\chi$PT expansion~\cite{Gasser:1983yg,Scherer:2005ri}.

The thermodynamic and low-energy properties of mesons at nonvanishing $\mu_I$ have been studied using the $\chi$PT in many different works~\cite{Son:2000xc,Kogut:2001id,Mammarella:2015pxa,Carignano:2016rvs,Carignano:2016lxe}. In particular, it has been confirmed that the pion condensed phase first discussed in~\cite{Migdal:1971cu,Sawyer:1972cq,Scalapino:1972fu, Migdal:1973zm,1972PhLA...41..129K} sets in at $\mu_I=m_\pi$, where $m_\pi$ is the pion mass.   Remarkably, $\chi$PT can also be used to study different gauge theories with isospin asymmetry, including 2 color QCD with different flavors~\cite{Kogut:1999iv,Kogut:2000ek, Hands:2000ei,Kogut:2001na, Brauner:2006dv,Braguta:2016cpw, Adhikari:2018kzh}

In the present paper we study  the multicomponent meson systems in which each  component is characterized by a global symmetry. In general, for each component, the spontaneous breaking of a global symmetry should lead to the formation of a superfluid. 
Multicomponent superfluids can be  realized in  He$^3$ - He$^4$ mixtures, see~\cite{Tuoriniemi2002,PhysRevB.85.134529} or in ultracold atoms experiments~\cite{CW,annabook,fallanibook,2014Sci...345.1035F,lvm2015}. In the compact star interior neutrons and protons are believed to simultaneously condense~\cite{Shapiro:1983du} and if deconfined quark matter is formed, {\color{black} the color-flavor locked phase~\cite{Alford:1998mk} supplemented by kaon condensation~\cite{Bedaque:2001je, Kaplan:2001qk} is a phase with two bosonic superfluids.}
Here we examine  the effect of  the possible intra-species interactions on multicomponent superfluidity. We focus on the meson condensed phase, employing   the $\chi$PT framework for deriving the relevant low-energy  Lagrangian.   We identify two very different types of interactions: those that  lock the two global symmetries and those that do not lock them. Remarkably,  at the leading order (LO) in  $\chi$PT, only the former type of interactions are possible. This kind of interaction  is not typically considered in ultracold  gases, because in these systems the number of particles of the two species are separately conserved.  In our work, we assume that this  not the   case  and we find that the strength of the locking term plays a prominent role.  Increasing the locking, we obtain that the transition to the broken phase is favored. Moreover,  for sufficiently large couplings the system becomes unstable. Analyzing the dispersion laws of the low-energy degrees of freedom, we find that the instability can be interpreted as a transition toward an inhomogeneous phase.  

Including the next-to-leading order (NLO) $\chi$PT corrections, it is possible to include interactions that do not lock the two chiral groups.  This type of interaction is akin to the one typically discussed in  ultracold atoms systems and indeed in this  case we obtain results similar to those of multicomponent Bose gas~\cite{Haber:2015exa}. 

The present paper is organized as follows. In Sec.~\ref{sec:one_gas} we report known results for meson systems in $\chi$PT. This is useful to fix the notation and for comparison with the multicomponent meson system. In Sec.~\ref{sec:two_gases} we generalize the $\chi$PT Lagrangian to two meson gases, introducing the leading interaction terms. In Sec.~\ref{sec:CL} we analyze the effect of one of the  possible  interaction term leading to chiral locking. In Sec.~\ref{sec:ICR} we consider the  $\chi$PT term  that does not lock the two chiral groups. We conclude in Sec.~\ref{sec:conclusions}. A number of results are collected in the Appendices. In the Appendix \ref{sec:appendixLow}, we report the low-energy excitations of a single-component pion gas. In the Appendix \ref{sec:appendixLHY}, we discuss the low-energy  corrections to the mean-field thermodynamic quantities arising from the vacuum energy of the Bogolyubov modes.

\section{Single meson gas} 
\label{sec:one_gas}
The $\chi$PT description {\color{black} of the single meson gas} is based on the global symmetries
\be G=SU(N_f)_L\times SU(N_f)_R\,,\ee 
 of massless QCD, with $N_f$ the number of flavors. 
The meson fields are collected in the $\Sigma$ field, transforming under $G$ as
\be
\Sigma \to  L \Sigma R^\dagger\,, 
\ee
where $L \in SU(N_f)_L$ and $R \in SU(N_f)_R$. The leading  $ {\cal O}(p^2)$  $\chi$PT  Lorenz-invariant Lagrangian~\cite{Kogut:2001id,Gasser:1983yg,Scherer:2002tk}  is given by  
\begin{align}
\label{eq:Lagrangian-single}
{\cal L} = \frac{f_\pi^2}{4} \text{Tr} (D_\nu \Sigma D^\nu \Sigma^\dagger) + \text{Tr} (M \Sigma^\dagger + M^\dagger \Sigma)\,,
\end{align}
where the mass matrix, $M$, and the so-called pion decay constant, $f_\pi$,  are the low energy constants (LECs) that cannot be fixed by the symmetry group $G$ and must be determined in some other way. The $\chi$PT Lagrangian is constructed assuming that  the mass term does not break the global symmetries, thus  that $M$  transforms as $\Sigma$. Then, the locking of the chiral rotations to the vector  $SU(N_f)_V$ group is induced by the vev of $M$, see for example the discussion in~\cite{Georgi:1985kw,Scherer:2002tk}. 

The covariant derivative in Eq.~\eqref{eq:Lagrangian-single} allows us to take into account the coupling of the meson fields with  the gauge fields and/or with   external currents and/or the effect of different {chemical} potentials~\cite{Gasser:1983yg, Leutwyler:1993iq,Leutwyler:1996er}. In the present work, we will only consider the effect of the isospin chemical potential and we will restrict the analysis to  pions, corresponding to the $N_f=2$ case. Thus, we consider the  covariant derivative 
\be
D_\nu \Sigma = \partial_\nu\Sigma - \frac{i}{2} \delta_{\nu 0} \mu_I [\sigma_3,\Sigma] \,,
\ee
where the { isospin chemical} potential, {\color{black} $\mu_I$}, is introduced as the time component of a vector field. Note that the covariant derivative does not include  the baryonic chemical potential, $\mu_B$, because  mesons  do not have a baryonic charge. A useful parameterization  is
\be\label{eq:single_sigma}
\Sigma = \cos \rho + i \hvf \cdot {\bm \sigma} \sin\rho \,,
\ee 
where the radial field, $\rho$,   and the unit vector field, $\hvf$, encode in a nontrivial way the three pion fields. By this parameterization, the LO $\chi$PT low-energy Lagrangian takes the form obtained in~\cite{Carignano:2016lxe} 
\begin{align}
\label{eq:lag-rho-phi}
{\cal L}=& \frac{f_\pi^2}{2} \left(\partial^\mu \rho\partial_\mu \rho + \sin^2\rho \; \partial^\mu \hf_i\partial_\mu\hf_i \right.\nonumber\\ &\left. -2 m_\pi \gamma \sin^2\rho  \; \epsilon_{3i k} \hf_i \partial_0 \hf_k \right) -V(\rho)\,, 
\end{align}
where
\be\label{eq:V}
V(\rho) = -f_\pi^2 m_\pi^2\left(\cos\rho + \frac{\gamma^2}{2}\sin^2\rho\right)\,,
\ee
is the potential and the control parameter is $\gamma = \mu_I/m_{\pi}$. For $|\gamma| >1$, the  pion condensed phase is favored~\cite{Migdal:1971cu,Sawyer:1972cq,Scalapino:1972fu, Migdal:1973zm,1972PhLA...41..129K,Migdal:1990vm,Son:2000xc,Kogut:2001id} and in the present parametrization it corresponds to a radial field vev, $\bar \rho$, satisfying 
 \be \cos\bar\rho =  \frac{1}{\gamma^2} \, .
 \ee
Therefore, in the broken phase the meson field vev is given by
\be\label{eq:condensate}
\bar \Sigma = \cos \bar\rho + i {\bm n} \cdot {\bm \sigma} \sin\bar\rho\,,
\ee 
where ${\bm n}$ is a unit vector associated  to the residual $O(2)$ symmetry of the vacuum. 
The  pressure and the isospin number density in the broken phase are respectively given by~\cite{Son:2000xc,Kogut:2001id,Carignano:2016rvs} 
\be
P = \frac{f_\pi^2 m_\pi^2}{2} \gamma^2 \left(1-\frac{1}{\gamma^2}\right)^2,\qquad  n_{I}= f_\pi^2 m_\pi \gamma \left(1-\frac{1}{\gamma^4}\right)\,, \label{eq:pressure-density}
\ee
leading to the ${\cal O}(p^2)$ equation of state~\cite{Carignano:2016rvs}
\be
\epsilon(P) = -P + 2\sqrt{P(2f_\pi^2 m_\pi^2+P)}  \label{eq:eq-state}\,.
\ee
Close to the phase transition point, $\gamma \gtrsim 1$,  the system is dilute and it is possible to expand {\color{black} the pressure $P$ and the energy density $\epsilon$, as a function of the isospin number density $n_I$}. If we define the adimensional {\color{black}isospin} density $n=n_I/(f_\pi^2 m_\pi)$, we can  expand the control parameter as    
\be\label{eq:expang}
\gamma = 1 + \frac{n}{4} + \frac{3 n^2}{32} + \frac{n^3}{32}  + {\cal O}(n^4)\,,
\ee
which is meaningful expansion for $n\ll 1$.  The pressure can then be expanded as follows
\be
P = \frac{n_I^2}{8 f_\pi^2} +\frac{n_I^3}{16 f_\pi^4 m_\pi^2} + {\cal O}(n_I^4)\,,
\label{eq:mfp}
\ee
where the leading term is the mean-field expression of the pressure of a boson system with coupling $g_0=1/4 f_\pi^2$. This is indeed the correct expression of  the coupling  close to the phase transition, see Eq.~\eqref{eq:g_expansion} and the discussion in the Appendix~\ref{sec:appendixLow}. The energy density is instead given by
\be
\epsilon = m_\pi n_I +  \frac{g_0 n_I^2}{2} + \frac{g_0^2 n_I^3}{2 m_\pi} + {\cal O}(n_I^4)\,,
\ee
which takes into account the energy associated to the mass of the pions.
Note that the above expressions are  obtained in the mean-field approximation, meaning that the low-energy fluctuations have not been included. Indeed,  the order $n_I^3$ corrections  are determined by   the  $\chi$PT Lagrangian and not by the contribution of the Bogolyubov modes. The vacuum contribution of the Bogolyubov modes  is considered in the Appendix~\ref{sec:appendixLHY}, and is much smaller than the {\color{black} leading mean-field contribution.} However, it can play an important role in a multicomponent gas, as we will see below.

\section{System of  two meson gases}
\label{sec:two_gases}

{\color{black} We now generalize the discussion of the previous Section to a  system with two mesonic gases.} In the second quantization formalism we assume that  {\color{black} two meson systems with densities $n_1$ and $n_2$} are described by the fields $\Sigma_1$ and $\Sigma_2$. As for the single meson gas discussed in the previous Section, we use the global symmetries for constructing the  $\chi$PT Lagrangian.
  
As a starting point we consider  the noninteracting case with  symmetry group \be\label{eq:G} G= G_1 \times G_2 \,,\ee
where \be\label{eq:Ga} G_a = \{SU(N_f)_L \times SU(N_f)_R \}_a  \qquad \text{with    } a=1,2  \ee is the chiral group of the $\Sigma_a$ field. {\color{black} For simplicity, we will mainly treat the system in which the two meson gases correspond to two fictitious pion systems, paving the way for the discussion of the simultaneous condensation of pions  and kaons.} In other words, we assume that in the noninteracting case the fields $\Sigma_1$ and $\Sigma_2$  transform independently under two chiral groups as 
\be
\Sigma_1 	\rightarrow L_1 \Sigma_1 R_1^\dagger \qquad \text{and}\qquad \Sigma_2 	\rightarrow L_2 \Sigma_2 R_2^\dagger\,,  \ee
where $L_a \in SU(2)_{L,a}$ and $R_a \in SU(2)_{R,a}$ with $a=1,2$.
The most general ${\cal O}(p^2)$  chiral  Lagrangian invariant under these symmetries is
\begin{align}
\label{eq:Lagrangian}
{\cal L} =& \frac{f_{1\pi}^2}4 \text{Tr} (D^1_\nu \Sigma_1 D^{1 \nu} \Sigma_1^\dagger) +   \frac{f_{2\pi}^2}4 \text{Tr} (D^2_\nu \Sigma_2 D^{2\nu} \Sigma_2^\dagger)\nonumber\\ & + 
\text{Tr} (\Sigma_1 M_1^\dagger + M_1 \Sigma_1^\dagger ) + \text{Tr} (\Sigma_2 M_2^\dagger + M_2 \Sigma_2^\dagger ) \,,
\end{align}
where $f_{1\pi}$ and $f_{2\pi}$, as well as  the matrices $M_1$ and $M_2$, are the low energy constants (LECs) of the system. As for a single meson system described by the Lagrangian in Eq.~\eqref{eq:Lagrangian-single}, we have constructed this Lagrangian assuming that  the mass terms do not break the global symmetries, which means that   $M_a$  transforms as $\Sigma_a$. The covariant derivative $D^a_\nu$ takes into account the interaction of the mesons of the $a$ system  with the external fields. If the  two meson systems have different { isospin chemical} potentials, $\mu_1$ and $\mu_2$, respectively,  this can be encoded in the two covariant derivatives 
\be\label{eq:covariantD}
D^\nu_a \Sigma_a = \partial^\nu\Sigma_a - \frac{i}{2} \delta_{\nu 0} \mu_a [\sigma_3,\Sigma_a] \,,
\ee
for $a=1,2$. 

We now  introduce the interaction between the two gases. Before doing that, 
let us first recall that under $G_a$ the covariant derivative transforms as the $\Sigma_a$ fields, that is 
\be
D^\mu_a \Sigma_a \to L_a D^\mu_a \Sigma_a R_a^\dagger\,, 
\ee
and therefore the two covariant derivatives are independently rotated. 
Let us now consider the possible interaction terms.  If we add to the noninteracting Lagrangian the term
\be
\label{eq:interaction_1}
{\cal L}_\text{int,1} = k \frac{f_{1\pi} f_{2\pi}}{2}  \text{Tr} (D^1_\nu \Sigma_1 D^{2 \nu} \Sigma_2^\dagger)\,, 
\ee 
it locks the two chiral groups, leaving only the diagonal  chiral rotation
\be
G_D= SU(N_f)_L \times SU(N_f)_R\,,
\ee
unbroken. In principle, the $k$ coefficient is a number that depends on the interaction strength between the two chiral fields and, as any LEC, it is independent of the { isospin chemical} potentials. 

Remarkably, the interaction Lagrangian in Eq.~\eqref{eq:interaction_1} is the only ${\cal O}(p^2)$ meaningful  coupling 
 leaving  the  $G_D$ group unbroken. One may think to add a Lagrangian term of the type
\be\label{eq:sis2}
\text{Tr} (\Sigma_1 \Sigma_2^\dagger)\,,
\ee
which indeed locks chiral rotations.   {\color{black} However, if  one of the two fields vanishes, from Eq. \eqref{eq:single_sigma} we have that  say $\Sigma_1 \equiv I$.} Then  the term in Eq.~\eqref{eq:sis2} acts as a mass term for the $\Sigma_2$ field,  breaking $G_2$ down to the vector subgroup. Therefore, this kind of term or any term of the type
 \be
\text{Tr} (\Sigma_1 \Sigma_2^\dagger)^n\,,
\ee
with $n>0$ is not allowed. For a similar reason the mass-like terms
 \be
\text{Tr} (M_1 \Sigma_1^\dagger  (\Sigma_2 \Sigma_1^\dagger)^n)\,,
\ee
are not allowed, unless $n=0$.

If one wants to preserve the $G$ group, then one has to consider  the  ${\cal O}(p^4)$ terms. {\color{black} At this order, there} are only two  derivative terms coupling the two meson systems   that do not lock the two chiral groups:
\begin{align} \label{eq:interaction_2}
{\cal L}_\text{int,2}=& \tilde L_1  \text{Tr} (D^1_\mu \Sigma_1 D^{1 \mu} \Sigma_1^\dagger)  \text{Tr} (D^2_\nu \Sigma_2 D^{2 \nu} \Sigma_2^\dagger)\nonumber \\  &+  \tilde L_2  \text{Tr} (D^1_\mu \Sigma_1 D^{1 \nu} \Sigma_1^\dagger)  \text{Tr} (D^2_\mu \Sigma_2 D^{2 \nu} \Sigma_2^\dagger)\,,
\end{align}
where $\tilde L_1$ and $\tilde L_2$ are two LECs analogous to the standard  $L_1$ and $L_2$ of ${\cal O}(p^4)$ $\chi$PT~\cite{Scherer:2002tk}.  When including these contributions, one should consistently include the standard  ${\cal O}(p^4)$ chiral terms, as well. However, as was shown in~\cite{Carignano:2016lxe}, the effect of the standard NLO terms on the thermodynamic properties of the system is extremely small and can be accounted for by a renormalization of the LO LECs. 

As an  aside, we note that in principle one may consider more complicated intra-species interaction terms, like 
\be
\label{eq:interaction_1new}
{\cal L}_\text{int} \propto   k^{\mu\nu}   \text{Tr} (D^1_\mu \Sigma_1 D^{2}_\nu \Sigma_2^\dagger)\,, 
\ee 
with $k^{\mu\nu}$ a Lorentz tensor and a  $G$ singlet. This kind of interaction term somehow generalizes Eq.~\eqref{eq:interaction_1} and Eq.~\eqref{eq:interaction_2}, however it  is not obvious how to fix the values of  the $k^{\mu\nu}$ components in general.

In the following, we will discuss  the effect of the  interaction terms in Eq.~\eqref{eq:interaction_1} and in Eq.~\eqref{eq:interaction_2}, separately,  focusing on the pion system.

\section{Chiral locking}
\label{sec:CL}
{To gain insight on the system described by Eqs.~\eqref{eq:Lagrangian} and  \eqref{eq:interaction_1}, let us first assume that 
we are making a partition of an ensemble of undistinguishable pions}, dividing it in two (interacting) subsets, in such a way that the $\Sigma_1$ field describes the pions of the first  subset and  $\Sigma_2$ field the pions of the second subset. Let us first focus on the kinetic terms at vanishing { isospin chemical} potentials. Since the pions are indistinguishable, one may naively think that the most general ${\cal O}(p^2)$ Lagrangian is

\begin{align}\label{eq:Lk}
{\cal L}=&  \frac{f_{\pi}^2}4 \text{Tr} (\partial_\nu \Sigma_1 \partial^{\nu} \Sigma_1^\dagger) +   \frac{f_{\pi}^2}4 \text{Tr} (\partial_\nu \Sigma_2 \partial^{\nu} \Sigma_2^\dagger)\nonumber \\ &+k \frac{f_{\pi}^2}{2}  \text{Tr} (\partial_\nu \Sigma_1 \partial^{\nu} \Sigma_2^\dagger)\,,
\end{align}
where  the first term, respectively the second term, describes the propagation and self-interactions of the fields of the subset $1$, respectively $2$. The third term mixes the two fields and induces the locking between the two subsets. If it were absent, that is for $k=0$, there would be no interactions between the two sets.  

For subsets  made of identical particles there must exist a way of reshuffling them. Since   $\Sigma_1\Sigma_1^\dagger+\Sigma_2\Sigma_2^\dagger=2$, any reshuffling can only correspond to a  rotation
\begin{align}\label{eq:rotation}
\Sigma_1 & \to  \cos\theta \, \hat \Sigma_1 + \sin\theta\,  \hat \Sigma_2\,, \nonumber \\
\Sigma_2 & \to  -  \sin\theta\,  \hat \Sigma_1 + \cos\theta\,   \hat \Sigma_2\,, 
\end{align}
transforming the Lagrangian  in Eq.~\eqref{eq:Lk} in 
\begin{align}\label{eq:L3}
{\cal L}=& + \frac{f_{\pi}^2}4 (1- k \sin (2 \theta)) \text{Tr} (\partial_\nu \Sigma_1 \partial^{\nu} \Sigma_1^\dagger) \nonumber  \\ & + \frac{f_{\pi}^2}4 (1+k \sin (2 \theta)) \text{Tr} (\partial_\nu \Sigma_2 \partial^{\nu} \Sigma_2^\dagger)\nonumber\\ &+k \frac{f_{\pi}^2}{2}  \cos(2 \theta)\text{Tr} (\partial_\nu \Sigma_1 \partial^{\nu} \Sigma_2^\dagger)\, .
\end{align}
{\color{black} To maintain} the Lagrangian invariant { we have to take $k=0$ or, more interestingly, $k  = 1$. Indeed, in the latter case}
\begin{align}\label{eq:lag_rotated}
{\cal L}=&  \frac{\hat f_{1\pi}^2}4 \text{Tr} (\partial_\nu \hat \Sigma_1 \partial^{\nu} \hat\Sigma_1^\dagger) +   \frac{\hat f_{2\pi}^2}4 \text{Tr} (\partial_\nu \hat\Sigma_2 \partial^{\nu}\hat \Sigma_2^\dagger)\nonumber\\ &+ \frac{\hat f_{1\pi}\hat f_{2\pi}}{2}  \text{Tr} (\partial_\nu\hat \Sigma_1 \partial^{\nu}\hat \Sigma_2^\dagger)\,,
\end{align}
where $\hat f_{1\pi}^2 = f_\pi^2 (1- \sin{2 \theta}) $,  $\hat f_{2\pi}^2 = f_\pi^2 (1+ \sin{2 \theta}) $, and therefore
 $\hat f_{1\pi} \hat f_{2\pi} = f_\pi^2 \cos{2 \theta}$. Note that one cannot identify $\hat f_{a\pi}$  with the pion decay constant of the pions in the subset $a$, because the  fields are mixed by the interaction terms. 
 
If one takes $k\neq 1$,   the $O(2)$ symmetry in Eq.~\eqref{eq:rotation} does not hold and the coefficient of the interaction term cannot be expressed as $\hat f_{1\pi}\hat f_{2\pi}$, meaning that if one makes the rotation, this term would depend on the rotation angle.
In the Lagrangian in Eq.~\eqref{eq:lag_rotated} it is possible to eliminate the dependence on the unphysical angle $\theta$ in the  quadratic terms  by writing
\be\label{eq:expansion}
\hat\Sigma_a= e^{ {i {\bm \sigma \cdot \bm{\hat \varphi_a}}/{\hat f_{a\pi}} }}\,,
\ee
which is a generalization of the standard nonlinear expression of the pion fields. Therefore, the expression in Eq. \eqref{eq:lag_rotated}, {\color{black} where $ k =1$ is set}, is the most general $\chi$PT Lagrangian  for two gases of  undistinguishable pions. We can easily generalize it to $N$ undistinguishable pion gas, writing
\begin{align}\label{eq:Lgeneral}
{\cal L}= \sum_{ab} \frac{f_a f_b}{4} \text{Tr} (\partial_\nu \Sigma_a \partial^{\nu} \Sigma_b^\dagger) \,,
\end{align}
where $f_{a}$ are a generalization of the pion decay constant. Note that  the propagating degrees of freedom are obtained by diagonalizing the quadratic Lagrangian. 

\begin{figure}[t!]
\centering
\includegraphics[width=0.45\textwidth]{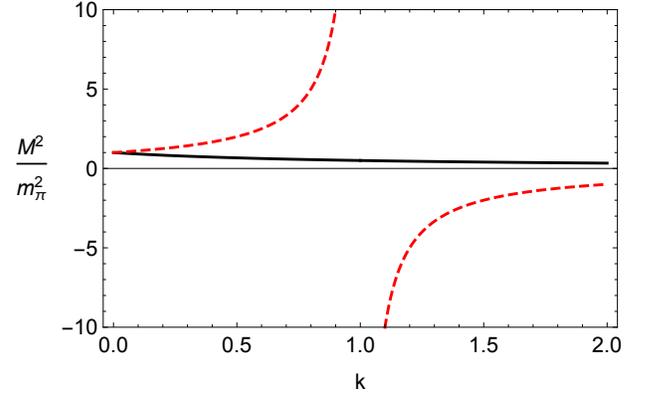}
\caption{Mass splitting induced by the  locking term in Eq.~\eqref{eq:interaction_1} for a two pion system. For simplicity we have assumed that the two gases have equal mass parameters, {\it i.e.} $m_{1\pi} =m_{2\pi}$. The  $k$  parameter indicates the strength of the intra-species locking, see Eq.~\eqref{eq:interaction_1}; $k=0$ corresponds to two noninteracting systems, while $k=1$ to a system of one single type of particles. For $k>1$ the system is unstable.}
\label{fig:splitting}
\end{figure}

Including the  mass terms, formally considering the vevs  of the fields $M_a$ in Eq.~\eqref{eq:Lagrangian}, we can write the total Lagrangian of the system as follows
\begin{align}\label{eq:Lagrangian_with_masses}
{\cal L}=&  +\frac{f_{1\pi}^2}4 \text{Tr} (\partial_\nu \Sigma_1 \partial^{\nu} \Sigma_1^\dagger) +   \frac{f_{2\pi}^2}4 \text{Tr} (\partial_\nu \Sigma_2 \partial^{\nu} \Sigma_2^\dagger)\nonumber \\ &+  \frac{f_{1\pi}^2 m_{1\pi}^2}4  \text{Tr} (\Sigma_1 + \Sigma_1^\dagger) +   \frac{f_{2\pi}^2 m_{2\pi}^2}4  \text{Tr} (\Sigma_2 + \Sigma_2^\dagger) \nonumber \\ &+k \frac{f_{1\pi}f_{2\pi}}{2}  \text{Tr} (\partial_\nu \Sigma_1 \partial^{\nu} \Sigma_2^\dagger)\,,
\end{align}
where we have assumed that the two fields have different  mass parameters, $m_{1\pi}$ and  $m_{2\pi}$. These parameters have to be interpreted as LECs for the coupled system and correspond to the pion masses only in the $k=0$ case.   The actual masses can be obtained by the dispersion laws
\be E_\pm^2= p^2 + M_\pm^2\,,\ee
where the masses are given by 
\be
M_\pm^2 = \frac{m_{1\pi}^2+m_{2\pi}^2\pm\sqrt{(m_{1\pi}^2-m_{2\pi}^2)^2+4 k^2 m_{1\pi}^2 m_{2\pi}^2 } }{2(1-k^2)}
\ee
for $k\neq 1$, and  equal to the ``reduced mass" 
\be\label{eq:dispersion}
M^2= \frac{m_{1\pi}^2 m_{2\pi}^2}{m_{1\pi}^2+ m_{2\pi}^2}\,,
\ee
for $k=1$. From the above expressions it is clear that the interaction term in Eq.~\eqref{eq:interaction_1} induces a  mass splitting. For clarity we report the behavior of the meson masses as a function of $k$ in Fig.~\ref{fig:splitting}.

We remind that $k=0$ corresponds to two noninteracting gases, while $k=1$ corresponds to two identical pion gases. For $k<1$, the mass splitting induced by the locking term is similar to the one induced by $\mu_I$ between the charged pions, see for example~\cite{Mammarella:2015pxa}. However, the system is unstable for $k>1$.  The instability  is signaled by the  divergent mass of one mode as $k \to 1^-$,  which becomes imaginary  for $k > 1$.  In the context of ultracold atoms physics,  where boson condensates are mostly considered, the latter feature is generally related to the appearance of spatially inhomogeneous phases, see e.g. \cite{baskin1975,smerzi2002} and references therein.
We stress, however, that here we are in the presence of a completely different instability.  Indeed, in ultracold atoms, the instability is triggered by a sufficiently large  coupling between the two systems~\cite{baskin1975,bohn1997,petrov2015} (a similar phenomenon is known also for fermions, called Stoner instability, see e.g. \cite{blundell}). Instead, in the present case, the locking plays the game: indeed, as $k$ varies, the repulsion from the locking term remains fixed and reads
\beq
V = -\sum_a f_{a\pi}^2 m_{a\pi}^2 \, .
\label{repnocond}
\eeq
In spite of this relevant difference and considering that the locked theory in Eq.~\eqref{eq:Lagrangian_with_masses} is quadratic, it is still quite natural to postulate that the same theory with $k >1 $ cannot exist with the two species coexisting in the same space domain. 

To elucidate the mechanism underlying the locking instability, and its possible resolution, let us consider  a simple system consisting of  two scalar bosons with a locking term
\begin{align}\label{eq:simplified}
{\cal L}=& {\cal L}_1 + {\cal L}_2 + {\cal L}_\text{int} = \frac{1}2 \partial_\mu \phi_1 \partial^\mu \phi_1 -  \frac{1}2 m_1^2 \phi_1^2\nonumber \\ &+  \frac{1}2 \partial_\mu \phi_2 \partial^\mu \phi_2 -\frac{1}2 m_2^2 \phi_2^2 + k\, \partial_\mu \phi_1 \partial^\mu \phi_2\,,
\end{align}
with a manifest discrete $Z_2 \times Z_2$ symmetry for $k=0$.  This symmetry corresponds to the transformations $\phi_1 \to - \phi_1 $ and $\phi_2 \to - \phi_2 $, separately. For $k \neq 0$ the two discrete symmetries are locked, with the only remaining $Z_2$ symmetry corresponding to $\phi_1 \to - \phi_1 $ and $\phi_2 \to - \phi_2 $, simultaneously.

This simple system becomes unstable for $k>1$, because one of the two eigenmodes has an imaginary mass. One possible solution of the instability corresponds to the realization of an inhomogeneous phase. Let us give an heuristic argument in favor of the inhomogeneous phase. If we assume that one component is realized in the volume $V_1$ and the other in the volume $V_2$, then the action can be written as
\begin{align}
{\cal S} &= \int d^4 x \, {\cal L} \approx  \int_{V_1} d^4 x \, {\cal L}_1 + \int_{V_2} d^4 x \,  {\cal L}_2  + \int_{S_{12}}  d^4 x \,  {\cal L} \nonumber \\ &= {\cal S}_1 + {\cal S}_2 + S_\text{interface}\,,
\end{align}
where ${\cal S}_a$ with $a=1,2$ are the actions of the free scalar fields. The effect of the interaction term is only relevant at the interface, ${S}_{12}$, of the two volumes. In other words, in the inhomogeneous phase   the interaction Lagrangian ${\cal L}_\text{int}$ has only support at the interface and  therefore the dispersion laws of the field $\phi_1$, respectively  $\phi_2$,  in the volumes $V_1$, respectively $V_2$, are not tachyonic.

\subsection{Two pion gases at different {\color{black} isospin chemical potentials}}

We now consider the effect of the isospin chemical potentials for the  two pion gases. Including {\color{black} them,}  the Lagrangian reads
\begin{align}
\label{eq:Lagrangian_su2}
{\cal L} =&+ \frac{f_{1\pi}^2}4 \text{Tr} (D^1_\nu \Sigma_1 D^{1 \nu} \Sigma_1^\dagger) +   \frac{f_{2\pi}^2}4 \text{Tr} (D^2_\nu \Sigma_2 D^{2\nu} \Sigma_2^\dagger)\nonumber\\ &  +
 \frac{f_{1\pi}^2 m_{1\pi}^2 }4 \, \text{Tr} (\Sigma_1 + \Sigma_1^\dagger ) +
 \frac{f_{1\pi}^2 m_{1\pi}^2 }4 \, \text{Tr} (\Sigma_2 + \Sigma_2^\dagger )  \nonumber\\ &+  k \frac{f_{1\pi}f_{2\pi}}2 \text{Tr} (D^1_\nu \Sigma_1 D^{2\nu} \Sigma_2^\dagger)\,,
\end{align}
where the covariant derivatives are given in Eq.~\eqref{eq:covariantD}.

Since the two fields can have  different vevs, we generalize Eq.~\eqref{eq:condensate} to
\be\label{eq:condensates}
\Sigma_a = \cos \rho_a + i {\bm n}_a \cdot {\bm \sigma} \sin\rho_i \qquad a=1,2\,,
\ee 
where  $\rho_a$ are the two  radial fields and ${\bm n}_a$ are two unit vectors. 
Upon substituting Eq.~\eqref{eq:condensates} in Eq.~\eqref{eq:Lagrangian_su2}, we obtain the tree-level potential
\begin{widetext}
\begin{align}
\label{eq:V_1}
V = -\sum_a f_{a\pi}^2 m_{a\pi}^2 \left( \cos\rho_a +\frac{\gamma_a^2}2 \sin^2\rho_a\right)  - k \,\bm{n_1} \cdot\bm{n_2} \,f_{1\pi }f_{2\pi }\mu_1 \mu_2 \sin\rho_1\sin\rho_2\,, 
\end{align}
\end{widetext}
where $\gamma_a=\mu_a/m_{a\pi}$ and the last term on the right hand side originates from the locking term, which explicitly breaks the $G$ symmetry  to the diagonal group, $G_D$ 
The interesting aspect is that the potential depends  on the relative angle between ${\bm n}_1$ and ${\bm n}_2$. In the ground state the two unit vectors are 
locked to be aligned, if the {isospin chemical} potentials have equal signs, or anti-aligned, if the {isospin chemical}  potentials have opposite signs.  We can clearly  restrict the analysis to the case in which both {isospin chemical} potentials are positive and aligned.  Since the vevs of the two fields are not independent but tend to align, it is clear that  the condensation of one field favors the condensation of the other;  we will discuss this effect in detail below. From the above expression it is also clear that the system has two NGBs for $k=0$, corresponding to the  two independent oscillations of the unit vectors, but only one  NGB  for $k\neq 0$,  corresponding to the locked oscillations of the two fields. The second mode is massive  and  corresponds to a  pseudo NGB.

\subsection{Phase diagram of the locked pion gases}
At the transition to the broken phase, where both gases condense, we can expand
\be
\cos\bar\rho_1 = 1 -\epsilon_1 \qquad \text{and} \qquad  \cos\bar\rho_1 = 1 -\epsilon_2 \,,
\ee
with $\epsilon_a\ll 1$. Upon replacing this expression in the stationary condition for the potential, we obtain
\begin{align}
\sqrt{\frac{\epsilon_2}{\epsilon_1}} = \frac{1-\gamma_1^2}{k \gamma_1\gamma_2}\qquad
\sqrt{\frac{\epsilon_1}{\epsilon_2}} = \frac{1-\gamma_2^2}{k \gamma_1\gamma_2}\,,
\end{align}
signaling that the condensation of one gas is deeply related to the condensation of the other:   as soon as, say, $\epsilon_1 >0$, it follows that $\epsilon_2>0$. The formation of one superfluid necessarily makes the other  gas superfluid by a simultaneous condensation (SCO) mechanism. 

\begin{figure}[th!]
\centering
\includegraphics[width=0.45\textwidth]{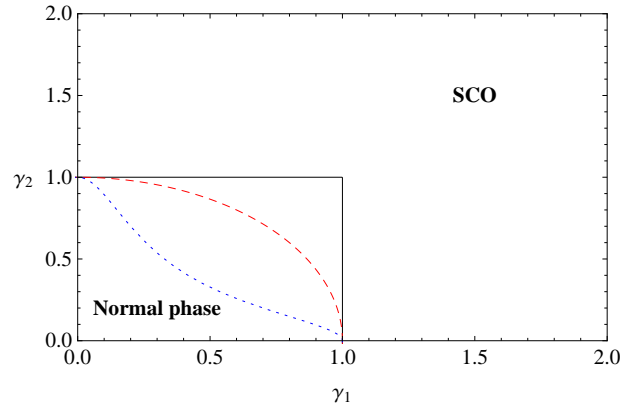}
\caption{Phase diagram for the locked two bosons gas system with interaction term in Eq.~\eqref{eq:interaction_1}.
The solid black line corresponds to $k=10^{-3}$; the dashed red line corresponds to $k=1$ and the dotted blue line corresponds to $k=5$. For every considered value of $k$, the broken phase is  the region outside the corresponding curve. It corresponds to a system in which there is the simultaneous condensation of both fluids  and is indicated with SCO. The only region where the SCO does not happen is along the axes, where $\gamma_1=0$ and $\gamma_2 >1$ or $\gamma_2=0$ and $\gamma_1 >1$; along these lines only one component is superfluid. The analysis of the low-energy excitations shows that for $k>1$ one of the low-energy modes becomes tachyonic, meaning that in this case the mean-field results reported in this figure are not valid.  }
\label{fig:phase_diagram1}
\end{figure}

Upon solving the above system of equations, we easily obtain that the SCO happens for
\be\label{eq:gamma12}
(k^2-1)\gamma_1^2 \gamma_2^2 + \gamma_1^2+\gamma_2^2=1\,,
\ee
corresponding to  the curve, ${\cal C}$ on the $(\gamma_1, \gamma_2) $ plane depicted in Fig.~\ref{fig:phase_diagram1} for various values of $k$.
The existence of this curve makes explicit that   the onset of one condensate induces the condensation of the other, a manifestation of the interaction between the two. A remarkable aspect is that the SCO  happens for any nonvanishing value of $k$. Clearly, the larger is $k$, the larger is the effect of one condensate on the other. Moreover, 
with increasing values of $k$, the normal phase region shrinks. To better understand this process, let us focus on  the $\gamma_1=\gamma_2=\gamma$ case. Since the two {isospin chemical} potentials are equal, it follows  has that $\bar\rho_1=\bar\rho_2=\bar\rho$,
\be
\cos{\bar\rho}= \frac{1}{\gamma^2(1+k)}\,,
\ee
and the transition happens for $\gamma^2=1/(k+1)$. Therefore, with increasing values of $k$, the transition to the SCO phase  happens at lower values of $\gamma$. One may naively think that  increasing  $k$ would lead to a system that becomes superfluid for arbitrary values of the isospin chemical potential. As we will see below, this is not the case, because for $k>1$ an instability in the low-energy spectrum is triggered.

In general, close to the transition curve, ${\cal C}$, one can expand the pressure as
\be
P = \frac{1}{2} L_{11} (\gamma_1-\bar\gamma_1)^2 +\frac{1}{2} L_{22} (\gamma_2-\bar\gamma_2)^2 +  L_{12} (\gamma_1-\bar\gamma_1) (\gamma_2-\bar\gamma_2)\,,
\ee
where $\bar\gamma_a \in {\cal C}$ and
\begin{align}
L_{ab} =\left. \frac{\partial^2 P}{\partial \mu_a \partial \mu_b}\right\vert_{\cal C}\,,
\end{align}
are the susceptibilities. 
Upon expressing the { isospin chemical} potential in terms of the number densities, we obtain
\be\label{eq:Pninj}
P = \sum_{ij} \frac{g_{ij}}{2} n_i n_j \,,\ee
where the coupling constants are given by
\be
g_{11} = \frac{L_{11}}D \qquad g_{22} = \frac{L_{22}}D \qquad g_{12} = -\frac{L_{12}}D\,,
\ee
where $D=L_{11} L_{22} - L_{12}^2$, {with  $L_{ab} > 0$}. 
{ It turns out that 
\be
g_{12} + \sqrt{g_{11} g_{22}} \geq 0 \,,
\label{eq:condstab}
\ee
 the equality corresponding to the case $\gamma_1 =\gamma_2=1/\sqrt{2}$. For non relativistic distinguishable and dilute superfluid bosons, the equality in Eq. \eqref{eq:condstab} corresponds to the stability threshold  against collapse or turn into an  inhomogeneous phase (depending on the sign of $g_{12}$) \cite{baskin1975,bohn1997,petrov2015}. By a similar reasoning, one could expect that, because of the relation in Eq.~\eqref{eq:condstab}, the two-pion locked system at nonvanishing isospin density is stable. More in detail,  the expression in   Eq.~\eqref{eq:condstab} relies on the mean-field approximation. {\color{black} Instead, in} condensed matter system it is known that the inclusion of the vacuum energy contribution of the Bogolyubov modes can only turn a collapsing system into an inhomogeneous one, {\color{black} made of droplets of coexisting gases}~\cite{petrov2015}. {\color{black} Anyway, in the present case the condition in Eq.~\eqref{eq:condstab} is not violated, the mean-field pressure is well defined, and the system could be expected to  be  homogeneous and stable.} However, for $k>1$, we found  that in the normal phase  there exists a tachyonic mode.  {\color{black}  It is therefore important to analyze the low-energy spectrum of the system to figure out what is the fate of the tachyonic mode in the SCO phase.}}

\subsection{Low-energy excitations}
The low-energy excitations of the multicomponent system can be determined studying the fluctuations of the radial component and of the Bogolyubov modes. We shall employ the same formalism developed in~\cite{Carignano:2016lxe} and briefly discussed in the Appendix~\ref{sec:appendixLow}, extending it to the two-component pion system.
\subsubsection{Radial excitations}
In the broken phase, the  system has two radial  excitations, $\chi_1$ and $\chi_2$,  corresponding to the fluctuations around the corresponding  vevs:
\be
\rho_a = \bar\rho_a + \chi_a \qquad \text{with  } a=1,2
\ee
where it is assumed that $\chi_a  \ll \bar\rho_a$. Upon substituting the above expression in Eq.~\eqref{eq:Lagrangian_su2} and restricting to the  quadratic order in the fields, we obtain the Lagrangian 
\begin{align}
{\cal L}_\chi= & \frac{1}{2} \partial_\mu \chi_1 \partial^\mu \chi_1 + \frac{1}{2} \partial_\mu \chi_2 \partial^\mu \chi_2+ c_{12}   \partial_\mu \chi_1 \partial^\mu \chi_2\nonumber\\ & - \frac{M_1^2}2 \chi_1^2 - \frac{M_2^2}2 \chi_2^2 + M_{12} \chi_1 \chi_2\,,
\end{align}
where 
\begin{align}
c_{12}&= \cos(\bar\rho_1 - \bar\rho_2)\qquad  s_{12}= \sin(\bar\rho_1 - \bar\rho_2)\nonumber\\
M_1^2&=m_\pi^2(\cos \bar\rho_1 - \gamma_1^2 \cos 2\bar\rho_1 + k\, \gamma_1\gamma_2\sin \bar\rho_1 \sin \bar\rho_2 )\nonumber\\
M_2^2&=m_\pi^2(\cos \bar\rho_2 - \gamma_2^2 \cos 2\bar\rho_2 + k\, \gamma_1\gamma_2\sin \bar\rho_1 \sin \bar\rho_2 )\nonumber\\
M_{12}&=k\, m_\pi^2 \gamma_1\gamma_2\cos \bar\rho_1 \cos \bar\rho_2\,.
\end{align}
The corresponding dispersion laws are given by \begin{widetext}
\be
E_\pm=\bm p^2 + \frac{c_{12} M_{12}+ (M_1^2+M_2^2)/2 \pm \sqrt{(c_{12} M_{12}+ (M_1^2+M_2^2)/2)^2+ s_{12}^2(M_{12}^2-M_1^2 M_2^2) } }{s_{12}^2}\,,
\ee
\end{widetext}
thus the two modes have non-negative masses and are stable for any value of $k$. 
On the  transition region to the BEC phase  $M_{12}=M_1 M_2$, and one of the radial modes becomes massless.  
The stability in the radial modes for any value of $k$ is clearly a manifestation of the result obtained in the previous Section,  that the pressure close to the transition region is positive defined.

\subsubsection{Bogolyubov modes}
Neglecting the radial excitations, thus taking $\rho_1 \equiv \bar\rho_1$ and $\rho_2 \equiv \bar\rho_2$, one has the following low-energy Lagrangian 
\begin{align}
{\cal L}_{\hf}= {\cal L}_1+ {\cal L}_2 + {\cal L}_{12}\,, 
\end{align}
where
\begin{align}
{\cal L}_1= \frac{f_\pi^2}2\left( \sin^2\bar\rho_1 \partial_\mu \hf^1_i \partial^\mu \hf^1_i + 2 \mu_1^2 \sin^2\bar\rho_i \epsilon_{3ik} \hf^1_i \partial_0  \hf^1_k \right)\,,
\end{align}
and with ${\cal L}_2$ given by a similar expression, while
\begin{align}
 {\cal L}_{12}=&k \, f_\pi^2 \left[\sin \bar\rho_1  \sin \bar\rho_2 ( \partial_\mu \bm \hf^1 \cdot \partial^\mu \bm \hf^2 + \mu_1 \mu_2  \bm \hf^1 \cdot  \bm \hf^2)\right.\nonumber\\
 & + \left. \epsilon_{3ik} ( \sin \bar\rho_1 \mu_2  \hf^2_i  \partial_0  \hf^1_k+  \sin \bar\rho_2 \mu_1 \hf^1_i  \partial_0  \hf^2_k ) \right]\,,
\end{align} 
stems from the locking term.
The unit vectors fields    $\hf_1$  and $\hf_2$ describe the two angular fluctuations of the condensates and can be parametrized as follows
\begin{align} \label{eq:def_hf}
\hf^1 = (\cos \alpha,\sin \alpha)  \qquad \text{and} \qquad \hf^2=  (\cos \theta,\sin \theta)\,,
\end{align}
which generalize the expression in Eq.~\eqref{eq:def_hf}. Upon substituting the above expression in the low-energy Lagrangian, we obtain 
\begin{widetext}
\be\label{eq:Lag_at}
 {\cal L}_{\hf}= \frac{f_\pi^2}2 \left[\sin^2 \bar\rho_1  \partial_\mu \alpha \partial^\mu \alpha + \sin^2 \bar\rho_2  \partial_\mu \theta \partial^\mu \theta + 2 k\sin \bar\rho_1  \sin \bar\rho_2 \cos(\alpha-\theta) (\partial_\mu \alpha \partial^\mu \theta+ \mu_1\mu_2)\right], 
\ee 
\end{widetext}
where we have not included the  terms
\be
f_\pi^2  \cos(\alpha-\theta)(\mu_2 \sin \bar\rho_1 \partial_0 \alpha + \mu_1 \sin \bar\rho_2 \partial_0 \theta)  
  \ee 
and 
\be
f_\pi^2 (\mu_1 \sin^2\bar\rho_1 \partial_0\alpha + \mu_2 \sin^2\bar\rho_2 \partial_0\theta) \,,
\ee
leading to interactions and total derivatives.
The Lagrangian in Eq.~\eqref{eq:Lag_at}  describes two coupled modes. We restrict to the case $\mu_1\mu_2 >0$; the other case can be treated in a similar way.  The potential term 
is minimized for $\alpha=\theta$, thus expanding in $(\alpha-\theta)$ and keeping only the quadratic terms, we obtain the dispersion laws
\begin{align}\label{eq:E12phonons}
E_1^2&= p^2 \nonumber  \\
E_2^2&= p^2 + \mu_1\mu_2 \frac{k(\sin^2\bar\rho_1 + \sin^2\bar\rho_2) + 2 k^2 \sin\bar\rho_1\sin\bar\rho_2}{1-k^2}\,,
\end{align}
corresponding to the {\color{black} massless NGB and the massive pseudo NGB}, respectively. The propagation velocity of the NGB is equal to $1$, however integrating out the radial oscillations would lead to a propagation velocity equal to the speed of sound, see~\cite{Carignano:2016lxe} and the discussion in the Appendix~\ref{sec:appendixLow}. For $k=0$  the mass of the pseudo NGB vanishes and thus  the system has two NGBs describing the independent fluctuations of  the two decoupled superfluids.

We notice that for $k \to1^-$ the mass of the pseudo-NGB diverges and  only one low-energy mode exists, which is consistent with the fact that for $k=1$  the system is equivalent  to a single superfluid. For $k>1$ the mass of the pseudo-NGB  becomes imaginary, signaling  an instability. This is the same instability we previously discussed in Fig~\ref{fig:splitting} in the unbroken phase. Thus, the unstable modes is still present in the SCO phase, now  appearing as a pseudo NGB with a   tachyonic mass.  {\color{black} The presence of this mode indicates that  the  mean-field approximation breaks down. Therefore, the expression of the pressure in Eq.~\eqref{eq:Pninj} is incorrect for $k>1$.} This result is discussed in more detail in the Appendix~\ref{sec:appendixLHY}, where it is shown that the beyond mean-field contributions are ill-defined for $k>1$.

\section{Independent chiral rotations}
\label{sec:ICR}

\begin{figure*}[t!]
\includegraphics[width=0.45\textwidth]{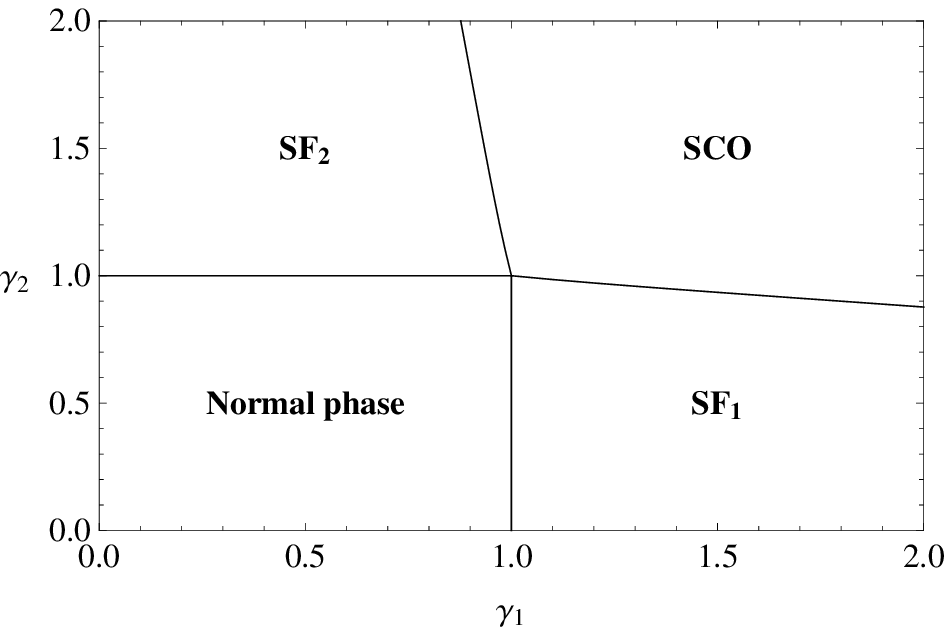}
\hspace{1.5cm}
\includegraphics[width=0.45\textwidth]{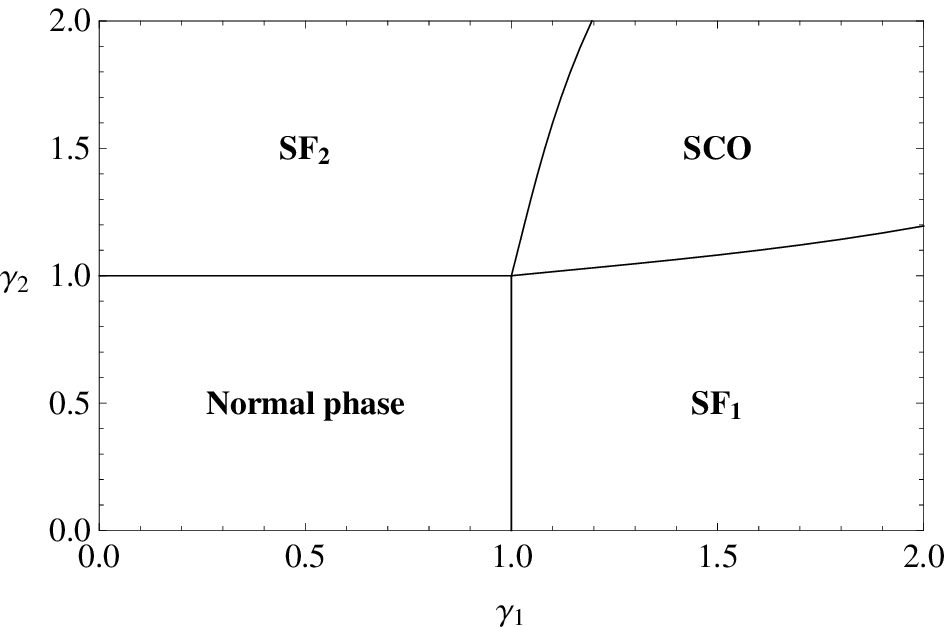}
\caption{Phase diagram for the coupled superfluid system with interaction term in Eq.~\eqref{eq:interaction_2}. Left {\color{black} panel}: case with $\tilde L_1+\tilde L_2 = + 10^{-2}$. Right {\color{black} panel}: $\tilde L_1+\tilde L_2 = -10^{-2}$. In both panels the phase with simultaneous condensation is indicated with SCO. The phases indicated with SF$_1$ and SF$_2$, correspond to the phases in which only one component is superfluid. Positive values of  $\tilde L_1+\tilde L_2$ favor the simultaneous condensation.}
\label{fig:phase_diagram2}
\end{figure*}

We now consider the  interaction terms that do not lock the two chiral groups.  Upon   expanding the Lagrangian given by Eqs. \eqref{eq:Lagrangian} and  \eqref{eq:interaction_2}, 
we obtain the potential 
\begin{widetext}
\begin{align}
\label{eq:V_2}
V = -\sum_a f_{a\pi}^2 m_{a\pi}^2 \left( \cos\rho_a +\frac{\gamma_a^2}2 \sin^2\rho_a\right)  -  4 (\tilde L_1+\tilde L_2) \, \mu_1^2 \mu_2^2  \sin^2\rho_1\sin^2\rho_2\,, \end{align}
\end{widetext}
where we have assumed  the two gases have unequal masses  and decay constant parameters. {\color{black} Unlike the locked case in the previous Sections},  now the tree-level potential is independent of the the relative orientation of the two condensates, indeed it does not depend on $\bm n_1 \cdot \bm n_2$. In other words, the potential does not break the degeneracy of the two vacua and the two condensates vectors $\bm n_1$ and $\bm n_2$ can independently rotate.  This is a manifestation of the fact that the interaction term does not lock the two chiral groups and thus the system has two NGBs. Considering $\tilde L_1+\tilde L_2 \sim 10^{-3}$, as typical for ${\cal O}(p^4)$ corrections (see for example~\cite{Scherer:2002tk})  the interaction  term has a small impact on the favored ground state. In particular, the onset of the simultaneous condensation is for $\gamma_1 \gtrsim 1 $ and $\gamma_2 \gtrsim 1$.  In the following we will consider $|\tilde L_1+\tilde L_2| = 10^{-2} - 10^{-3}$, {\color{black} also} taking into account possible negative values of {\color{black} $(\tilde L_1+\tilde L_2)$}.

In Fig.~\ref{fig:phase_diagram2}, we report the phase diagrams obtained with positive (left panel) and negative (right panel) values of  $\tilde L_1+\tilde L_2$. The behavior with the strength of the intra-species interaction is very similar to the one obtained for a coupled  two-fluid system in~\cite{Haber:2015exa}. The  $\tilde L_1+\tilde L_2$ parameter has the same effect on the phase diagram of the entrainment parameter of~\cite{Haber:2015exa}: a positive value of $\tilde L_1+\tilde L_2$ favors the SCO, while a negative value disfavors it.  In~\cite{Haber:2015exa} it was also discussed the instability generated by coupled superfluid flows. Although a similar phenomenon might emerge in our model,  we postpone its analysis to future work.

\begin{figure}[th!]
\centering
\includegraphics[width=0.45\textwidth]{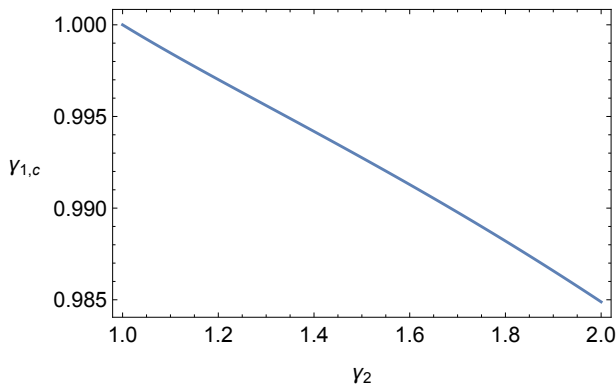}
\caption{Critical value for the condensation of the fluid $1$, once the fluid $2$ is in the superfluid phase, obtained by Eq.~\eqref{eq:independent_g1g2} for $\tilde L_1+\tilde L_2 = 10^{-3}$.}
\label{fig:independent_g1g2}
\end{figure}

In order to infer the effect of one superfluid on the other,  we consider the case in which  one of the two superfluids is formed, say the superfluid $2$, and we seek the critical value $\gamma_{1,c}$ for the onset of the condensation of the superfluid $1$. At the leading order in the intra-species interaction, we find that the condensation onset for the first species obeys the equation 
\be\label{eq:independent_g1g2}
\gamma_{1,c}^2=1-8 (\tilde L_1+\tilde L_2)\frac{\gamma_2^4-1}{\gamma_2^2}\,,
\ee
which is depicted in Fig.~\ref{fig:independent_g1g2} for $\tilde L_1+\tilde L_2 = 10^{-3}$.

In principle, for large values of $\gamma_2$ it suffices a small $\mu_1$ { isospin chemical} potential to drive the system $1$ in the condensed  phase. However,
for reasonable values of the NLO LECs,  the influence of one condensate on the other is extremely small. 
The low-energy spectrum in the broken phase consists of two NGBs which have a very small mixing. The system does not show any instability in the spectrum of the Bogolyubov modes.

\section{Conclusions}
\label{sec:conclusions}

We have discussed multicomponent meson superfluids in the $\chi$PT framework. We have derived the relevant $\chi$PT Lagrangian  restricting most of the analysis to the global symmetry group given in Eqs.~\eqref{eq:G} and \eqref{eq:Ga} with $N_f=2$,  corresponding to two fictitious pion gases with different masses and decay constants.  In the noninteracting case,  if one of the two {\color{black}  isospin chemical potentials} exceeds the corresponding pion mass the system becomes superfluid. Turning on the interactions the two condensates influence each other.  We have considered two possible interaction terms, one that locks the two chiral groups and one that does not lock them.

The Lagrangian term in  in Eq.~\eqref{eq:interaction_1}  leads to the tree-level potential in Eq.~\eqref{eq:V_1}, with the peculiar interaction term between the phases of the two condensates. Minimizing the potential we have obtained the phase diagram reported in Fig.~\ref{fig:phase_diagram1}. With increasing {\color{black} locking parameter $k$}, the region in which the simultaneous condensation is realized  becomes larger. It seems that one can arbitrarily shrink the normal phase region by increasing the value of $k$. However,  the locking turns one low-energy mode becomes in a pseudo NGB with dispersion law  given in Eq.~\eqref{eq:E12phonons}. For  $k>1$ the mass of the pseudo NGB becomes imaginary and therefore an instability is triggered. The unusual aspect is that even for $k>1$ the potential has a well defined minimum, indeed the low-energy radial excitations studied in Sec.~\ref{sec:radial} have a well-defined mass.
Since no other homogenous phase is energetically favored, this suggests that there exists an energetically favored inhomogeneous phase, {\color{black} where the two gases do not coexist any longer}. Though not rigorously  proved, this seems an educated guess, also because of  the analysis of the simplified model  discussed in Sect~\ref{sec:CL}. 
 It is not obvious to us that this inhomogeneous phase can be treated by a Ginzburg-Landau expansion~\cite{Ginzburg:1950sr},  or any other improved version~\cite{Carignano:2017meb},  because in these approaches one expects the appearence of an inhomogeneous phases when the mean-field analysis indicates a first-order phase transition. Instead, in the present case, the tree-level analysis does not show any phase transition or any instability: the only sign of an odd behavior is in the spectrum of the pseudo NGB mode. 
 
The Lagrangian term in Eq.~\eqref{eq:interaction_2}, which does not lock the two global symmetries, is also interesting, because it  induces a nontrivial interaction between the two condensates. However, in $\chi$PT this term can only arise at the NLO in the chiral expansion, thus we expect that it is strongly suppressed. The tree-level interaction potential  is reported in Eq.  \eqref{eq:V_2}: since it is independent of $\bm n_1$ and $\bm n_2$, it is clear that in this case the two condensates are free to oscillate and are not locked.  The low-energy modes consist of two NGBs which do not show any singular behavior. Upon minimizing the potential in Eq.~\eqref{eq:V_2} we obtain the phase diagrams reported in Fig.~\ref{fig:phase_diagram2}. 

The present work can be extended in different ways. As already anticipated, it paves the way for the discussion of a two-component system of pions and kaons. We plan to develop this study shortly. It would also be interesting to realize the locking instability in two-component ultracold atoms system.

\section{Acknowledgements}
{ The authors thank T. Macri, {\color{black} S. Paganelli,} L. Salasnich,  and A. Trombettoni for helpful discussions and B. B. Brandt for providing us the lattice data points shown in Fig.~\ref{fig:nI}.}
 
\appendix\section{Low-energy expansion}
In the following we recap and slightly extend the discussion of the low-energy modes of the pion condensed phase of~\cite{Mammarella:2015pxa}  and in~\cite{Carignano:2016lxe}, including higher order terms. 
\label{sec:appendixLow}


\subsection{Radial field}
\label{sec:radial}
Expanding  the radial field  around the stationary value as $\rho=\bar \rho +\chi$  and neglecting the angular fluctuations we obtain from Eq.~\eqref{eq:lag-rho-phi} 
\begin{align}\label{eq:L_chi}
{\cal L_\chi} =&  \frac{f_\pi^2}{2} \partial^\mu \chi\partial_\mu \chi+ f_\pi^2 m_\pi^2 \left(\frac{1-\gamma^4}{2 \gamma^2} \chi^2\right. \nonumber\\ &\left.-\frac{1}2\sqrt{\frac{\gamma^4-1}{\gamma^4}}\chi^3 + \frac{4 \gamma^4-7}{24 \gamma^2}\chi^4\right)\,,
\end{align}
where the ${\cal O} (\chi^5)$ terms and higher have been suppressed. It is convenient to rescale the field with $\chi \to \chi/f_\pi$ to put the kinetic term in the canonical form, obtaining
\begin{align}\label{eq:L_chi2}
{\cal L_\chi} =  \frac{1}{2} \partial^\mu \chi\partial_\mu \chi -    \frac{1}{2} m_\chi^2 \chi^2  -g_{3\chi}  \chi^3 +g_{4\chi}  \chi^4\,, 
\end{align}
where the mass  and self-couplings are given by \begin{align}\label{eq:mchi} m_\chi &= m_\pi \gamma \sin\bar\rho\,,\\
g_{3\chi} &= \frac{m_\pi^2 \sin\bar\rho}{2f_\pi}\,,\\
g_{4\chi} &= \frac{m_\pi^2}{f_\pi^2} \frac{4 \gamma^4-7}{24 \gamma^2}\,.
\end{align}
We notice that  the only nonvanishing term at the phase transition point is the one proportional to $\chi^4$. Actually, it can be easily proven that any term proportional to $\chi^{2n +1}$ vanishes at $\gamma=1$, because in the unbroken phase the system is symmetric for $\rho \to -\rho$. Close to the phase transition point, the radial fluctuations can be considered as a self-interacting system of bosons with vanishing mass and cubic interaction but nonvanishing quartic interaction. This Lagrangian for the radial fluctuation is valid in the whole broken phase. For the angular field the situation looks different.


\subsection{Bogolyubov mode}
The Lagrangian of the angular field is given by
\be\label{eq:L_phikin} {\cal L}= \frac{\hat f_\pi^2}{2}  \partial^\mu \hf_i\partial_\mu\hf_i  \qquad \text{for   } i=1,2
\ee
with $\hat f_\pi = f_\pi \sin\bar\rho$ playing the role of an effective decay constant. Since   $\hf$ is a unit vector, we can parameterize it by a {\color{black} Bogolyubov mode $\alpha$ as follows:   }
\be\label{eq:def_hf}
\hf_1= \cos \alpha \qquad \hf_2= \sin \alpha
\ee
leading to 
\be\label{eq:L_alpha} {\cal L}= \frac{\hat f_\pi^2}{2}  \partial^\mu\alpha \partial_\mu\alpha\,,
\ee
which is the Lagrangian of a free scalar field, $\alpha$. It can be cast in the canonical form by $\alpha \to \alpha/\hat f_\pi$. The Bogolyubov field can only feel the medium effect by the interactions with the $\chi$ field, as will be discussed below.
We note that the NLO chiral terms would be proportional to higher powers of momentum, therefore  this is the relevant Lagrangian only for $p^2/\hat f_\pi^2 \ll 1$. For this reason, this low-energy expansion is not valid close to the phase transition point, corresponding to $\gamma =1$, where  $\hat f_\pi$  vanishes and thus all the terms of the effective Lagrangian are equally important.  Since the momentum scale is dictated by the temperature of the system, one has  to consider the  $T/\hat f_\pi \ll 1$ case. 

\subsection{Mixed terms and dispersion laws}
The mixed terms can be obtained from the interaction terms in Eq.~\eqref{eq:lag-rho-phi} and considering that upon substituting Eq.~\eqref{eq:def_hf} we have the compact expression
\be
\epsilon_{3i k} \hf_i \partial_0 \hf_k  = \partial_0 \alpha\,,
\ee
in terms of the Bogolyubov field $\alpha$. Thus, up to the fourth order in the fields, the mixed interaction terms are
\begin{align}
\label{eq:mixed}
{\cal L}_I^{\chi \alpha} =&-g_{2,1} \chi\partial_0\alpha+  g_{3,1}\chi^2\partial_0 \alpha + g_{3,2}\chi \partial_{\mu}\alpha \partial^{\mu} \alpha
\\ &+ g_{4,1} \chi^3 \partial_0\alpha  + g_{4,2} \chi^2 \partial_{\mu} \alpha \partial^{\mu} \alpha\,,
\end{align}
with the  couplings  given by:
\begin{align}
g_{2,1}&= \frac{2 m_\pi}{\gamma} \qquad g_{3,1}= \frac{\gamma^4-2}{\gamma^3 \hat f_\pi} m_\pi\,,\\
g_{3,2}&= \frac{1}{\gamma^2 \hat f_\pi} \qquad g_{4,1}= \frac{4 m_\pi}{3 f_\pi^2 \gamma}\,,\\
g_{4,2}&= \frac{2-\gamma^4}{2 \gamma^4 \hat f_\pi^2}  \,,
\end{align}
where the first {\color{black} subscript} indicates the total number of fields and the second one the number of $\alpha$ fields.

The quadratic Lagrangian can be written as
\be
{\cal L}= \frac{1}{2} \partial^\mu \chi\partial_\mu \chi -    \frac{1}{2} m_\chi^2 \chi^2 +  \frac{1}2  \partial_{\mu} \alpha \, \partial^{\mu} \alpha  -g_{2,1} \chi\partial_0\alpha\,,
\ee
where  the mixing term allows oscillations between the radial and the angular fields. Integrating out the radial fluctuations one obtains the massless,  phonon-like, dispersion law
\begin{align}\label{eq:disp_phonon}
E_\text{ph}&= c_s p\,,
\end{align}
where
\begin{align}\label{eq:cs}
c_s & =\sqrt{\frac{m^2_\chi}{m^2_\chi+g_2^2}}= \sqrt{\frac{\gamma^4-1}{\gamma^4+3}}
\end{align}
describes the pressure oscillations propagating at the sound speed.

Alternatively, one can diagonalize the quadratic Lagrangian, obtaining the dispersion laws
\begin{align}
E_\pm &= \sqrt{p^2+\frac{m_\text{eff}^2}{2}\pm  \sqrt{\left( \frac{m_\text{eff}^2}{2}\right)^2 + g_2^2 p^2}}\,,
\end{align}
where 
\begin{align}
m_\text{eff}^2&= m^2_\chi+g_2^2=m_\pi^2\frac{\gamma^4+3}{\gamma^2}\,,
\end{align}
which agree with the expressions reported in~\cite{Mammarella:2015pxa}. In conclusion, the  low-energy modes of a single-component pion gas correspond to a NGB with dispersion law  in Eq.~\eqref{eq:disp_phonon} (in the limit $p/m_\chi \to 0$)  and to a massive mode with mass  $m_\text{eff}$.

\section{LHY correction}
\label{sec:appendixLHY}
Close to the phase transition  to the broken phase, the pressure of the single-component pion gas can be approximated with the expression in Eq.~\eqref{eq:pressure-density}. Therefore, $\chi$PT analysis gives a correction  to the mean-field value proportional to $n^3$. However, {\color{black} in the context of condensed matter physics, an additional contribution, due to the vacuum energy of the NGBs, is known to play an important role in certain regimes. This contribution} is known as the Lee-Huang-Yang  (LHY) term, first evaluated for a hard sphere Bose gas in~\cite{Lee:1957zzb}.  The LHY term is proportional to $n^{5/2}$ and is the leading correction to the mean-field results, {\color{black}  close to the phase transition point.}

For a single-component pion gas, one can easily obtain the LHY correction using the mapping developed in~\cite{Carignano:2016rvs} between the condensed pion gas in $\chi$PT and the Gross-Pitaevskii Hamiltonian 
\be\label{eq:hamiltonian}
{\cal H}_\text{GP} =  \psi^* \frac{\nabla^2}{2 M} \psi -\frac{g}2 \vert \psi^* \psi \vert^2\,,
\ee
where $M=\mu_I$, and 
\be\label{eq:g_expansion}
g =\frac{4 \gamma^2-1}{12 f_\pi^2 \gamma^2} = g_0\left(1 + \frac{n}6\right) + {\cal O}(n^3) \,,
\ee
where $g_0=1/(4 f_\pi^2)$ is the coupling constant at the phase transition point. The LHY correction to the pressure close to the phase transition point is given by
\be\label{eq:ELHY_1}
\epsilon_\text{GP,LHY} = \frac{M^{3/2}}{15 \pi^2}  (g \, n_I)^{5/2} \propto m_\pi^4 \, n^{5/2} \,,
\ee
with the particular dependence on $n$ indicating  that this is a nontrivial effect {\color{black} beyond mean field}.   The LHY  contribution is the first one in the series expansion $n a^3$, where $a= g M/(4 \pi)$ is the s-wave scattering length. Close to the transition point and using the values of the coupling constant and of the mass of the GP expansion, we find that  $n a^3 \ll 1$ {\color{black} that means the diluiteness condition} for any $\gamma \in [1,2]$.  However, the evaluation of the LHY term by  Eq.~\eqref{eq:ELHY_1}  assumes that the GP expansion is reliable, {\color{black} implying} that $1 \leq \gamma \ll 2$. 
\begin{figure}[t!]
\centering
\includegraphics[width=0.45\textwidth]{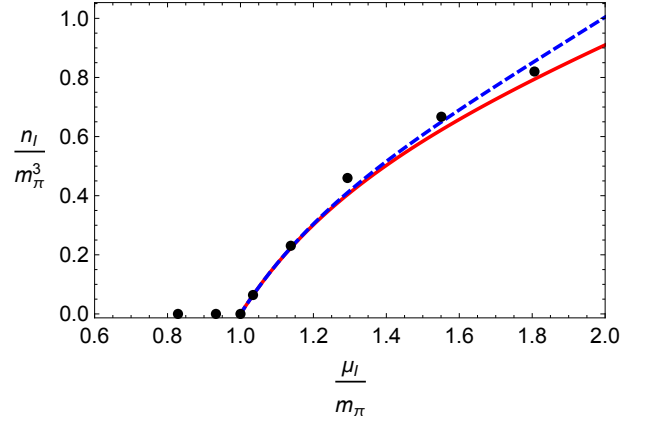}
\caption{Isospin number density as a function of the { isospin chemical} potential for a single-component pion gas. The solid red line corresponds to the LO $\chi$PT result. The dashed blue line is obtained adding the LHY contribution. The dots correspond to the lattice results of~\cite{Brandt:2017oyy, Brandt:2018bwq, Brandt:2018wkp}.}
\label{fig:nI}
\end{figure}

For a general evaluation of he  LHY correction in the $\chi$PT context,  we consider  the vacuum contribution of the NGBs
\be\label{eq:ELHY_2}
\epsilon_\text{LHY} \propto \frac{1}{2 \pi^2} \int_0^\Lambda d p \, p^2    E_\text{ph}\,,
\ee
where $E_\text{ph} = c_s p$ is the dispersion law of the NGBs obtained integrating out the radial fluctuations, see Eq.~\eqref{eq:disp_phonon}.
The hard cutoff, $\Lambda$, takes into account that the NGBs describe the low-energy fluctuations below the mass scale, $m_\chi$, of the radial field, see Eq.~\eqref{eq:mchi}. Taking for simplicity  $\Lambda = m_\chi$, 
considering the expression of the speed of sound in Eq.~\eqref{eq:cs}, and that, close to the phase transition, $
\gamma \approx 1 + n/4$, see Eq.~\eqref{eq:expang}, we find 
\be
\epsilon_\text{LHY} \propto m_\pi^4 n^{5/2}\,,
\ee
in agreement with Eq.~\eqref{eq:ELHY_1}. 
In Fig.~\ref{fig:nI} we compare the isospin number density evaluated in $\chi$PT (solid red line), with that obtained including the LHY correction (dashed blue line), as well as with  the results of the lattice simulations of Refs.~\cite{Brandt:2017oyy, Brandt:2018bwq, Brandt:2018wkp}  using the same value of their pion mass, $m_\pi = 135$ MeV, and of the pion decay constant, $f_\pi = 133/\sqrt{2}$.

The  $\chi$PT results systematically underestimate the number density. With the inclusion of the LHY term the agreement slightly  improves.  It follows that the   $\chi$PT + LHY pressure is always larger than the $\chi$PT one. However, the difference between the two is extremely  small. 

Generalizing the previous discussion to the two-component pion gases with the interaction term in Eq.~\eqref{eq:interaction_1}, it is clear that there are two relevant low-energy contributions. One from the NGB, and one from the pseudo NGB. Since the latter becomes tachyonic for $k>1$, {\color{black} the LHY} contribution is ill-defined.  Again, this is a signal that the mean-field approximation breaks down  for $k>1$, and thus the evaluation of the pressure  of the system given by the expression in Eq.~\eqref{eq:Pninj} is incorrect.

%

\end{document}